\documentclass[9pt,twocolumn,twoside]{osajnl}

\journal{ol} 

\setboolean{shortarticle}{true}

\title{Anomalous reflection at the interface of binary synthetic photonic lattices}

\author[1]{Zhiqing Zhang}
\author[1]{Yanan Dai}
\author[2]{Zengrun Wen}
\author[1]{Zhenjuan Liu}
\author[1]{Haohao Wang}
\author[2]{Yuanmei Gao}
\author[3]{Yanlong Shen}
\author[1*]{Xinyuan Qi}

\affil[1]{School of Physics, Northwest University, 710127 Xi'an, China}
\affil[2]{College of Physics and Electronics, Shandong Normal University, Jinan, 250014, China}
\affil[3]{State Key Laboratory of Laser Interaction with Matter, Northwest Institute of Nuclear Technology, Xi’an Shaanxi 710024, China}

\affil[*]{Corresponding author: qixycn@nwu.edu.cn}

\begin{abstract}
	We construct a binary synthetic photonic lattice theoretically with an effective magnetic field by projecting two fiber loops' light intensity and adjusting the phase distribution precisely. By tuning the phase modulator, wave vector, and propagation constant of an effective waveguide, the interface's transmittance could be manipulated. Further light dynamics show that the light pulse can achieve total reflection without diffraction and exchanges the light energy in two optical fiber loops completely when phase distribution and wave vector meet certain conditions. Our study may provide a new way to realize optical switches in optical interconnection and optical communication.
\end{abstract}

\setboolean{displaycopyright}{true}

\begin{document}
	
	\maketitle

	\label{sec:examples}
	
	The propagation and evolution of waves in discrete systems are of paramount importance for understanding various fundamental physical phenomena, including slow light~\cite{Luis_Slow_2021} and topological waves~\cite{Ablowitz_Solitons_2020}  in photonic lattices, acoustic spin transport~\cite{Long_Realization_2020}, and localization dynamics~\cite{Zhong_Photon_2020, Maczewsky_Synthesizing_2020} et al. In the field of optics, discrete systems generally refer to systems with discontinuous refractive index distribution, such as photonic crystal~\cite{Prudencio_First_2020}, photonic lattices~\cite{Ablowitz_Solitons_2020, Porras_Topological_2019}, and optical waveguide array~\cite{Li_An_2016, Vassholz_Observation_2021}. Synthetic photonic lattice (SPL), as an optical Galton board~\cite{Bouwmeester_Optical_1999}, is discrete in two scales: trip numbers of average length and length difference~\cite{Miri_Optical_2012}. In recent years, the SPL has become a multifunctional platform for studying many experimental phenomena due to its simple structure, low production cost, flexibility, and easy manipulation. A large number of important optical phenomena have been demonstrated in SPL, e.g., the Parity-Time (PT)-symmetric Talbot effect~\cite{Wang_symmetric_2018}, 2D light walk~\cite{Muniz_Collapse_2019}, and Bloch oscillation~\cite{Wimmer_Observation_2015}. Lately, the SPL was also employed to realize the light localization, for instance, localized modes at effective gauge field interface~\cite{pankov_observation_2019}, asymmetric localization in anti-PT SPL~\cite{dai_asymmetric_2020}, skin effect in non-Hermitian SPL~\cite{PhysRevB.97.121401, Weidemann_Topological_2020}, and topological Floquet interface states~\cite{Bisianov_Topological_2020}. These works expand the physical connotation of SPL.
	
	Reflection, as a kind of wave dynamics, has been investigated in numerous systems, such as spin waves that can be split into two different modes in magnetic materials~\cite{Tomosato_Bi_2020},  magnetic reflection incident upon a step-like variation of an artificial vector potential in atomic flux lattices~\cite{An_Direct_2017}, beam reflection at the interface of two two-dimensional (2D) arrays of evanescently coupled waveguides~\cite{Cohen_Generalized_2020}, and nonlinearity-dependent reflection in a sawtooth photonic lattice with defects~\cite{ji_nonlinearity-dependent_2018}. Nevertheless, in SPL, the observation of the reflection phenomenon is still remarkably absent.
	
	This paper proposes a binary synthetic photonic lattice (BSPL) with an effective magnetic field by dynamically adjusting the asymmetric phase distribution. Due to the difference in the structure of binary synthetic photonic lattice, the interface is produced, which leads to the phenomena of transmission and reflection. Afterward, the Gaussian beam's transmittance $T$ in the binary system is analyzed, and the result shows that $T$ will change accordingly as long as the phase $\Phi_0$, propagation constant $\Omega$, and wave vector $k$ are changed. Further study shows that the total reflection can be achieved and the energy in two fiber loops can be exchanged exactly when $(\Phi_0,\Omega,k,) = (\pi,\pi/2,n\pi)$ ($n \in Integer$). Compared with the previous work, we focus on studying the optical transmission at the interface and systematically analyze the total reflection in BSPL. Our work provides new ideas for realizing binary structures and has an important impact on the study of photodynamics. At the same time, it has potential applications in optical devices.
	\begin{figure}[h]
		\centering
		\includegraphics[width=8.5cm,keepaspectratio]{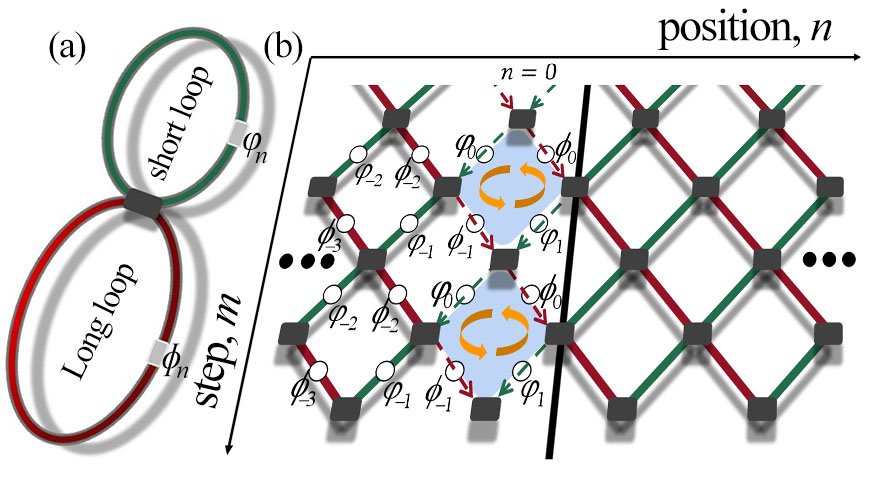}
		\caption{Time-multiplexed dynamics of a binary synthetica photonic lattices. (a) two coupled fiber loops of different lengths and corresponding phase distribution $\phi_n$ and $\varphi_n$; (b) spatial structure of the synthetic lattices. The shadow area indicates one plaquette.}
		\label{Fig.~1}
	\end{figure}
	
	First, we constructed two optical fiber loops with a length difference of $\Delta L $ connected by a directional coupler, as shown in Fig.~\ref{Fig.~1}(a). The time delay corresponding to the light beam's transmission in the two fiber loops is $2\Delta T$ = $\Delta L/c_F$, where $c_F$ is the speed of light in the fiber. When the optical pulse sequence is injected into one of the loops, it is split into two smaller pulse signals at the coupler and continues to propagate along with the two loops, respectively. By projecting the propagation of light pulse along the long or short loop as the $n+1$ and $n-1$ of the lateral position, and the number/time of the interference at the coupler as the longitudinal position $m$, an SPL is constructed [Fig.~\ref{Fig.~1}(b)]. Once the phase modulator in one loop is tuned dynamically, the SPL phase distribution will also be changed. As shown in Fig.~\ref{Fig.~1}[b], we construct a BSPL by setting the phase distribution asymmetrically: there is no phase adjustment in the long/short loop for $n>0$ and the position-relevant phases $\phi_n$, and $\varphi_n$ are tuned by phase modulators in the long and short loops, respectively, for $n \leq 0$ [see Fig.~\ref{Fig.~1}(a)]. $\phi_n$ and $\varphi_n$ can be expressed as
	
	\begin{equation}
	\begin{aligned}
	\phi_{n}=0, (n\in Integer)\\
	\varphi_{n}=\begin{cases}
	0, & \text {$(n>0)$}\\
	\varphi. & \text {$(n\leq 0)$} 
	\end{cases}
	\end{aligned}
	\label{eq:01}   
	\end{equation}
	
	The net phase circumfluence intensity $\Phi_{n}$ satisfies $\Phi_{n}=\varphi_n+\phi_{n-1} -\varphi_{n+1}-\phi_n$ [see the circle arrow in Fig.~\ref{Fig.~1}(b)]. Inserting Eq.~\ref{eq:01} into $\Phi_{n}$, the  phase circumfluence $\Phi_{0}=\varphi$ holds only at the plaquettes of $n=0$, while $\Phi_{n}=0$ for the other positions. Therefore, the effective magnetic field $B_{eff}$ arises from the phase circumfluence in one plaquette could be given by the relation $B_{eff} = \frac{1}{a^2} \int \vec{A}_{eff}\cdot d\vec{l} = \frac{\Phi_n}{a^2}$~\cite{PhysRevLett.108.153901}, where $a^2$ is the area of one plaquette, $\vec{A}_{eff}$ is the effective magnetic potential. This means that the magnetic field came into being once the net phase circumfluence intensity $\Phi_{n}$ in the BSPL exists ($\Phi_{n} \neq 0$). Considering the A-B effects for photons~\cite{PhysRevLett.108.153901}, we reasonably believe that some interesting light transmission behavior may occur at the interface of BSPL.
	
	To study the light pulse dynamics in SPL, we use complex amplitude $p_{n}^{m}$ and $q_{n}^{m}$ to represent the light field in long and short loops, respectively. Then the amplitude evolution in the sound pressure level can be described by the following discrete equation~\cite{dai_asymmetric_2020},
	
	\begin{equation}
	\begin{array}{l}
	p_n^{m+1}=[\cos(\alpha)p_{n+1}^m+i\sin(\alpha)q_{n+1}^m)]e^{-i\varphi_n},\\
	q_n^{m+1}=[\cos(\alpha)q_{n-1}^m+i\sin(\alpha)p_{n-1}^m)]e^{-i\phi_n},
	\end{array}
	\label{eq:02}   
	\end{equation}
	where $\alpha$ is the coupling coefficient, $\phi_n$ and $\varphi_n$ are the position-related phases for $n \leq 0$ [see Eq.~\ref{eq:01}].
	
	Consider the asymmetric structure of our BSPL, we pay special attention to the light transmission across the interface [as the black line shown in Fig.~\ref{Fig.~1}(b)]. Assuming that stationary methods with trial solutions of Eq.~\ref{eq:02} have the form of,
	
	\begin{equation}
	(p_{n},q_{n})^\mathrm{T}=exp(-i \Omega m)(P_{n},Q_{n})^\mathrm{T},
	\label{eq:03}   
	\end{equation}	
	where $P_{n}$ and $Q_{n}$ describe the $m$-independent amplitudes in the long and short loops, respectively. $\Omega$ is propagation constant in $m$ direction. One can obtain the following iteration relations Eq.~(\ref{eq:04}) by taking the trail solutions Eq.~(\ref{eq:03}) into Eq.~(\ref{eq:02}),   
	\begin{equation}
	P_{n-2}	= [e^{i(\Omega-\varphi_{n-2})} + e^{i(\varphi_{n-1}-\varphi_{n-2} -\Omega)}]\sec(\alpha)  P_{n-1} -e^{-i\varphi_{n-2}} P_n.
	\label{eq:04}	
	\end{equation}	
	
	If the light pulse is incident with an angle, a trial solution $P_n$ can be described with incident wave $R_0 e^{ikn}$,  reflected wave $R e^{-ikn}$ and transmitted wave $T_0 e^{ikn}$ as shown in Eq.(\ref{eq:05}),    
	\begin{equation}
	P_{n}=\left\{\begin{array}{ll}
	R_{0} e^{i k n} + R e^{-i k n}, &  (n \leq 0) \\
	T_{0} e^{i k n}, & (n > 0)
	\end{array}\right.
	\label{eq:05}	    
	\end{equation}
	where $k$ is the wave vector, $R_{0}, R$, and $T_{0}$ are the incident, reflected, and transmitted amplitudes, respectively. They can be calculated by the Eq.(\ref{eq:05}),
	\begin{equation}
	\begin{array}{l}
	R_{0}=\frac{\left(P_{0}e^{2i k}-P_{-1} e^{i k}\right)}{\left(e^{2 i k}-1\right)}, \\
	R=\frac{\left(P_{-1}e^{i k}-P_{0}\right)}{\left(e^{2 i k}-1\right)}.
	\end{array}
	\label{eq:06}
	\end{equation}
	
	Obviously, the incident and reflected amplitude $R_0$ and $R$ are function of $P_{0}$ and $P_{-1}$. Defining the transmittance of light pulse as $T= \frac{\left | T_0 \right |^2}{\left | R_0 \right |^2}$. Setting $\alpha=\pi/4$, the transmittance $T$ of light incident from left to right can be denoted as
	\begin{equation}
	T(\Phi_0, \Omega, k)=\left |\frac{(1-e^{-2ik})e^{i\Phi_0} }{1+(\varepsilon-e^{i(k-\Phi_0)})(e^{ik}-\lambda )e^{i\Phi_0}} \right | ^2,
	\label{eq:07}
	\end{equation}
	where 
	\begin{equation}
	\begin{array}{l}
	\varepsilon =2\sqrt{2}\cos(\Omega)e^{-i\Phi_0},\\
	\lambda =\sqrt{2}[e^{i(\Omega-\Phi_0)}+e^{-i\Omega}].\\
	\end{array}
	\label{eq:08}	
	\end{equation}
	
	From Eq.\ref{eq:07} and Eq.\ref{eq:08}, it is shown that the transmittance $T$ is dependent on the propagation constant $\Omega$, the phase circumfluence $\Phi_0$ at the position $n=0$, and the wave vector $k$. Cosidering that the system is closed, so the transmittance $T \leq 1$ and one can obtain the following relation,  
	\begin{equation}
		\begin{array}{l}
			2 \sqrt{2}[3+2\cos(2\Omega)]\cos(\frac{\Phi_{0}}{2}-\Omega) \sin(\frac{\Phi_{0}}{2}) +2\cos(\Phi_{0}-4\Omega) \\ +
			\cos(\Phi_{0})+ 8 \cos(\frac{\Phi_{0}}{2}-2 \Omega) \cos(\frac{\Phi_{0}}{2})+ 5\geq 0,
		\end{array}
		\label{eq:09}	
	\end{equation}
	here, $k=\pi/2$. Obviously, reasonable transmittance $T$ can only be obtained when Eq.~\ref{eq:09} is satisfied.
	
	\begin{figure}[ht]
		\centering{
			\includegraphics[width=8.5cm,keepaspectratio]{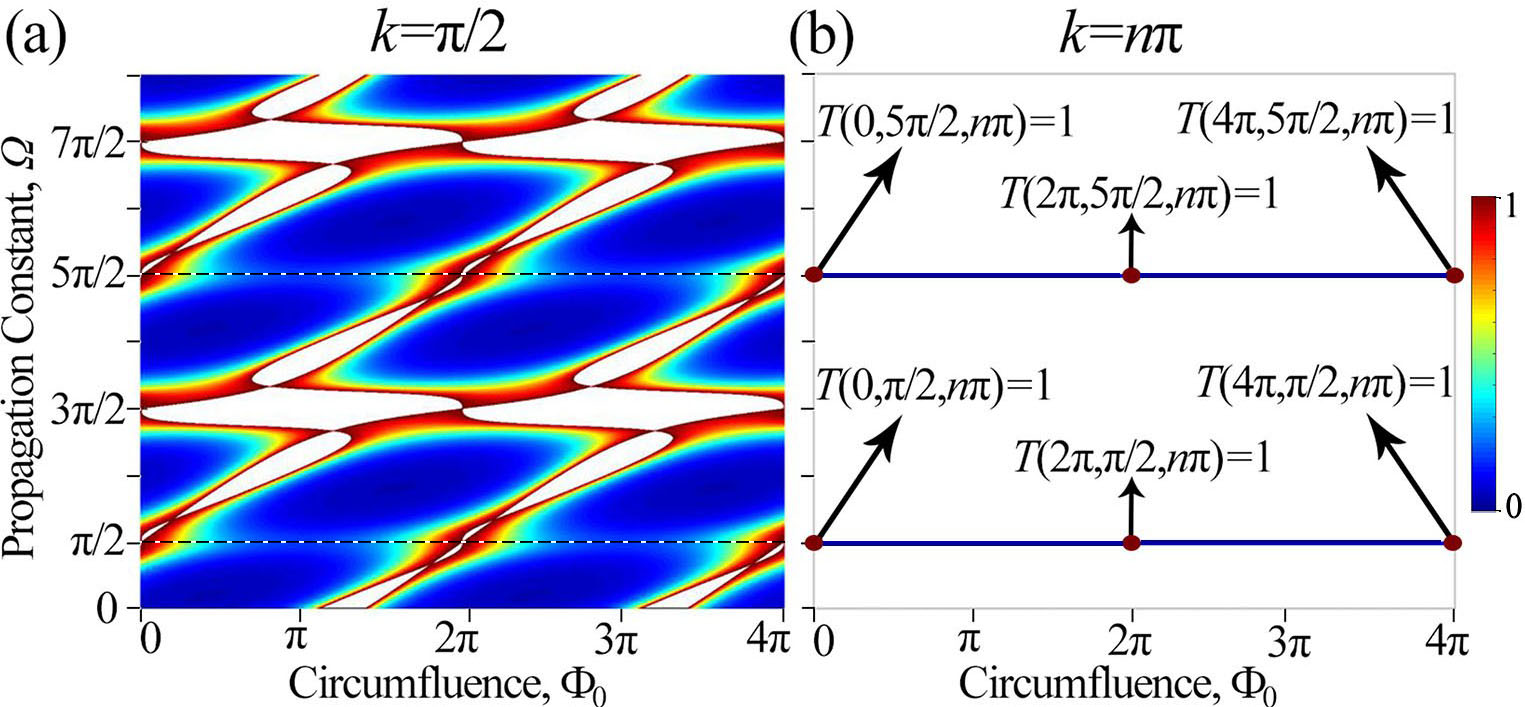}}
		\caption{A map of transmittance  $T$ as a function of phase circumfluence $\Phi_{0}$ and propagation constant $\Omega$ for $k=\pi/2$ (a), and $k=n\pi$ (b).}
		\label{Fig.~2}
	\end{figure}
	
	Figure.~\ref{Fig.~2} shows the map of transmittance $T$ versus $\Omega$ and $\Phi_{0}$ for the light beam with wave vector of $k = \pi/2$. The transmittance $T$ reaches its maximum value of 1 when $\Omega -\frac{2}{3}\Phi_0 = \frac{\pi}{2} +2n\pi$ [see Fig.~\ref{Fig.~2}(a)]; while $T$ has nonzero values, $T(0,\pi/2,n\pi)=1$, $T(2\pi, \pi/2,n\pi)=1$ and $T(4\pi, \pi/2,n\pi)=1$ when $\Omega = \pi/2$ [see the magenta dots in Fig.~\ref{Fig.~2}(b)]. Apparently, the light transmittance $T$ could be manipulated periodically by tuning the phase circumfluence $\Phi_0$ when $\Omega=\pi/2  +2n\pi$, and $T$ will be undefined when Eq.~\ref{eq:09} is not satisfited [see the white color in Fig.~\ref{Fig.~2}(a)]. Therefore, the following is discussed to reveal the relationship between $T$ and the phase circumfluence $\Phi_0$, wave vector $k$ when $\Omega = \pi/2$. 
	
\begin{figure}[ht]
		\centering{
			\includegraphics[width=8.5cm,keepaspectratio]{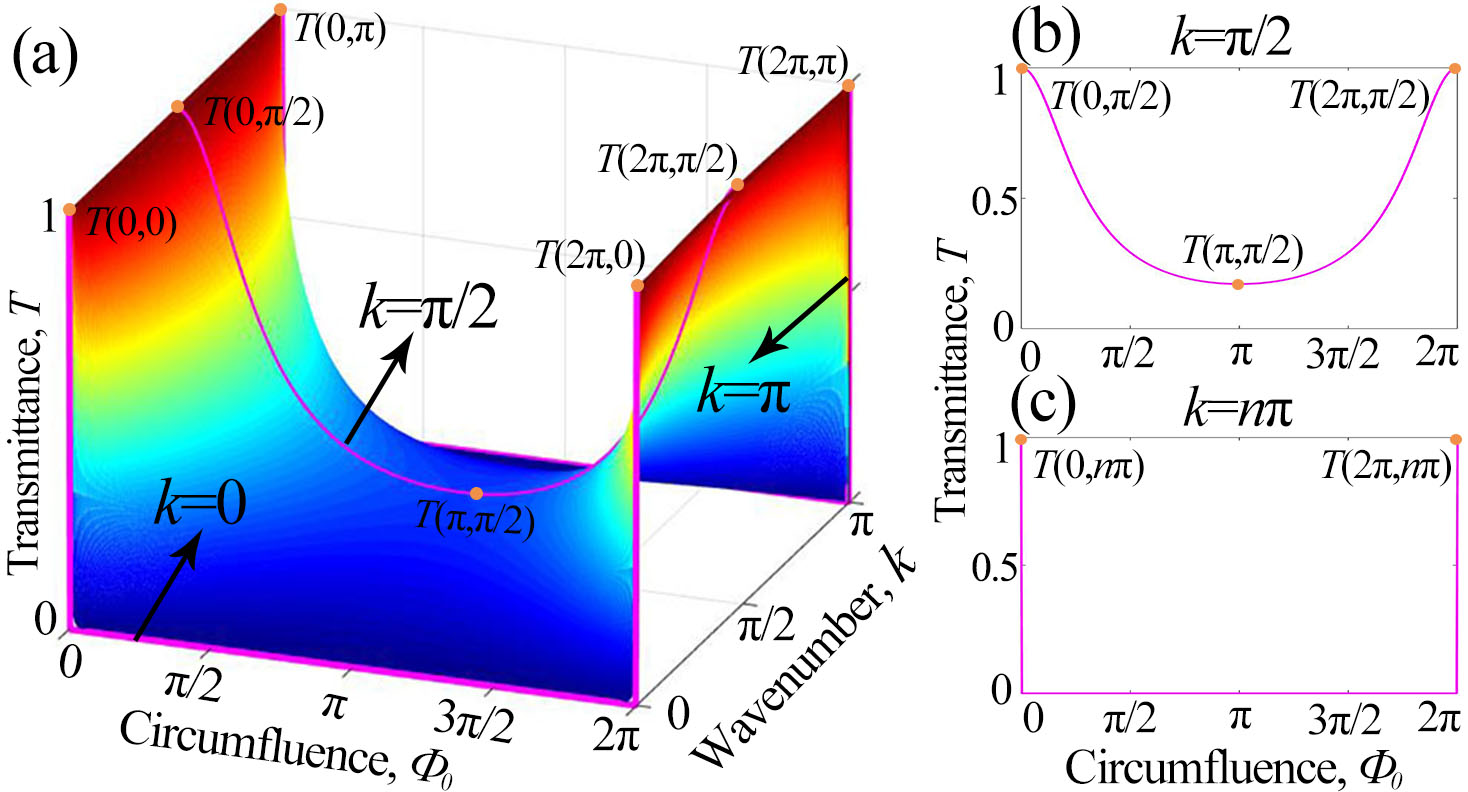}}
		\caption{The distribution pattern of transmittance  $T$ as a function of circumfluence $\Phi_0$ and wave vector $k$ for $\Omega = \pi/2$. (a), three-dimensional pattern in a period. (b) and (c), the curves of $T$ versus $\Phi_0$ for wave vectors of $k=\pi/2$ and $k=n\pi$, respectively.}
		\label{Fig.~3}
	\end{figure}

	As shown in Fig.~\ref{Fig.~3} (a), a saddle-shaped transmittance map of $T$  versus phase circumfluence $\Phi_0$ and wave vector $k$ in a period is demonstrated, here the propagation constant $\Omega = \pi/2$. For the cases of $k=\pi/2$,  $T$ is close to a parabolic-like curve [see the pink lines in Fig.~\ref{Fig.~3}(a) and (b)]. For this situation ($k=\pi/2$), $T(\Phi_0, k)$ has some special values $T(0, \pi/2)=1$, $T(2\pi, \pi/2)=1$ and $T(\pi, \pi/2)=0.17$. For the cases of $k=n\pi$, transmission $T$ reaches its maximum at $\Phi_0=0$ and its minimum for the other situations [see the purple lines in Fig.~\ref{Fig.~3}(a) and (c)]; Interestingly enough, such a kind of abrupt variation for transmittance directly from 1 to 0 indicates that the beam can be trapped in the left, and there is no energy transmission through the interface of our BSPL when $k=0$ and $k=\pi$ [see Fig.~\ref{Fig.~3}(a) and (c)].     
	
	To further understand the transmission characteristics at the interface of BSPL, we calculated the group dispersion relation of the left structure. Firstly, a Floquet-Bloch discrete plane wave ansatz is used as the solution of Eq.\ref{eq:02}, which reads
	
	\begin{equation}
	\left(
	\begin{array}{c}
	p_n^m\\
	q_n^m
	\end{array}
	\right) =\left(
	\begin{array}{c}
	P_n\\
	Q_n
	\end{array}
	\right)e^{i k n}e^{i \Omega m},
	\label{eq:10}
	\end{equation}
	where $\Omega$ is the propagation constant, $k$ is wave vector. By inserting Eq.\ref{eq:10} into Eq.\ref{eq:02}, the dispersion relation of the band structure can be obtained in the form of
	
	\begin{equation}
	\sqrt{2}\cos{(\Omega+\Phi_0/2)}-\cos{(k-\Phi_0/2)}=0,
	\label{eq:11}
	\end{equation}
	here, $\Phi_0 =\varphi$ is considered. Simultaneously, $\Omega$ is only related to $\Phi_0$ and wave vector $k$. 
	
	Considering the definition of group velocity dispersion $ D={\frac{\mathrm{d}^2\Omega}{\mathrm{d} k^2}}_{|k_0}$, $D$ has the following form of Eq.~\ref{eq:12} by solving the second-order differential of the band dispersion given by Eq.~\ref{eq:11}, 
	
	\begin{equation}
	D=\frac{\sqrt{2}\sin{(k-\Omega-\Phi_0)}}{-1+\cos{(2\Omega+\Phi_0)}}.
	\label{eq:12}
	\end{equation}
	
	Especially, $D=0$ means that the light beam is diffractionless during its transmission according to group velocity dispersion's physical connotation. A set of solutions can be obtained by solving Eq.~\ref{eq:11} for $D=0$, $(\Phi_0,\Omega,k)=(\pi,\pi/2, n\pi)$. Together, one can find that the transmittance $T=0$ with this set of solutions. In other words, the light beam will be diffractionless during the propagation and then be reflected completely at the interface of BSPL.
	
	\begin{figure*}[!htp]
		\centering
		\includegraphics[width=16cm,keepaspectratio]{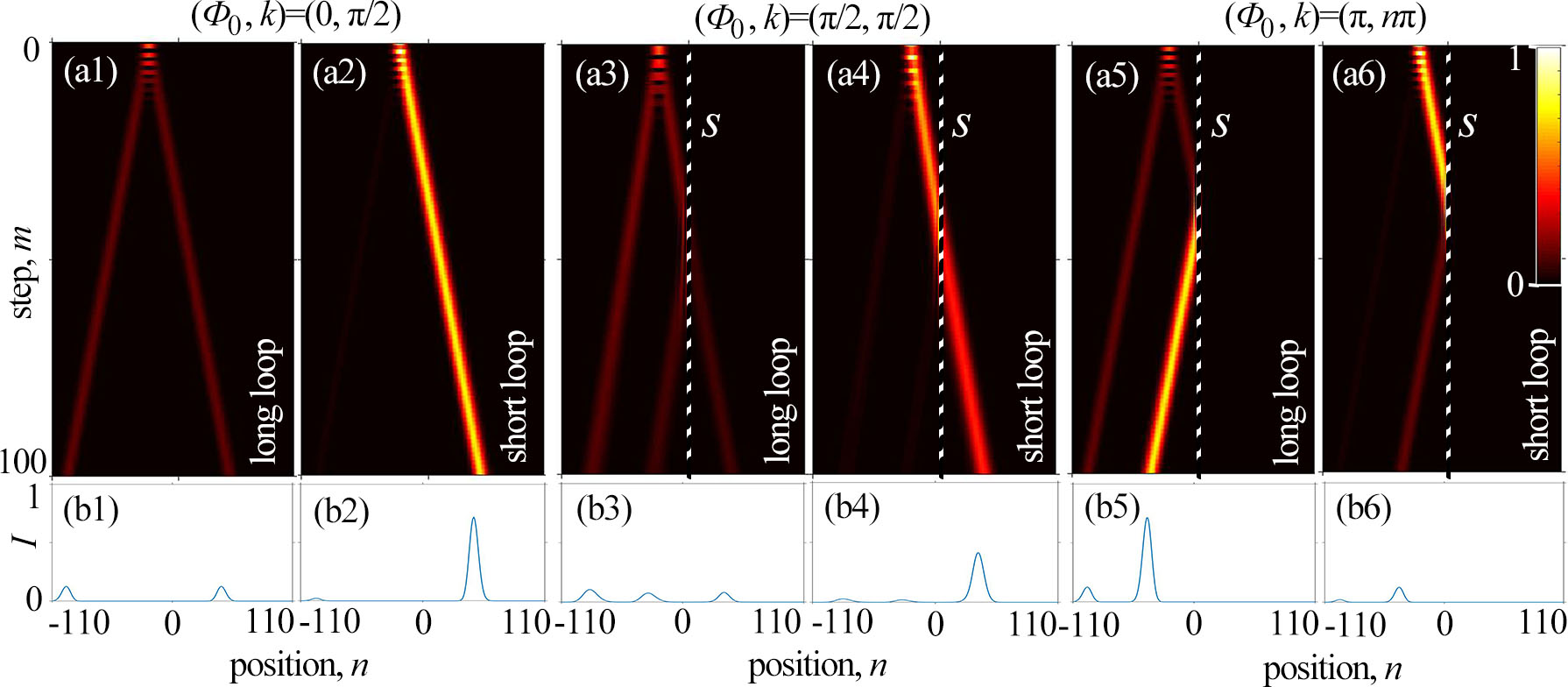}
		\caption{Numerically computed evolution of a Gaussian beam in the BSPL incident from a short loop. The top row, evolution of light intensity. The bottom row, the intensity profiles at the exit facet of the lattices. $I$ and the dashed lines $S$ indicate the light intensity and interface, respectively.}
		\label{Fig.~4}
	\end{figure*}
	To verify the above analysis, the light transmission of a Gaussian beam with an initial profile of $p_{n}^{0}=e^{(-n/\bigtriangleup)^2}e^{ikn}$ is investigated. The FWHM of a Gaussian beam is set to be $2\bigtriangleup=8$. For comparison, a Gaussian beam with initial conditions of $(\Phi_0, k) = (0,\pi/2)$ is injected into the short loop at the beginning, most of the light diffraction-free propagation in the short loop except that very few parts of light energy is scattered into the long loop and evolved into a bifurcated diffractionless light pattern, as shown in Fig.~\ref{Fig.~4}(a1) and (a2). The transmittance in the short loop is around $72.62\%$, which is very close to the result in Fig.~\ref{Fig.~3}(b). When the light beam is incident with conditions of $(\Phi_0, k) = (\pi/2,\pi/2)$, it is seen that around $52.95\%$ of the input energy will keep its initial shape and direction while the rest will be coupled in the long loop. Especially, one lobe of the light beam in the long loop will be reflected at the interface of BSPL [see the middle lobe in Fig.~\ref{Fig.~4}(a3)], while the transmitted light decreases in the short loop, as shown in Fig.~\ref{Fig.~4}(a4). This indicates there exists energy exchange at the interface of BSPL arising from the phase circumfluence $\Phi_0$. Further, one can see the anomalously total reflection at the interface of BSPL when the Gaussian beam is incident with conditions of $(\Phi_0, k) = (\pi,n\pi)$ [see Fig.~\ref{Fig.~4}(a5) and (a6), (b5) and (b6)]. Unlike the conventional reflection and the broadband anomalous reflection in meta-surface~\cite{Pu2013Broadband}, in our anomalous reflection, the reflected light energy in short and long loops originates exactly from the long and short loops, respectively. So, in other words, light energy in the short loop exchanges completely with one lobe in the long loop. Therefore, there is no light energy penetrate through the interface of BSPL. The transmittance $T(\Phi_0,\Omega,k)=T(\pi,\pi/2, n\pi)=0$, this agrees well with the theoretical analysis as shown in Fig.~\ref{Fig.~3}(c). Note that the decrease of calculated transmittance could be attributed to the scattering at the lattice's entrance facet.  
	
	An applicable experiment project to achieve this anomalous reflection could be performed with two fiber loops. The basic principle of the feasible experimental setup is inspired by~\cite{Feng_Single_2014, Yao_Edge_2018}. Two standard single-mode fibers @ wavelength $\lambda=1550$nm and average length $L$ are coupled by a $2\times 2$ coupler with a power ratio of $50/50$. The light pulse with a width of about tens nanosecond orders is injected into any of the loops (a short loop is selected in this paper). Then, the pulse intensity must change when part of the light beam is coupled to another loop at the coupler, which is detected by the photodiode. Simultaneously, with the help of an optical isolator to ensure that the propagation of pulses in any loops always remains unchanged along the initial direction, an erbium-doped fiber amplifier is used to compensate for the optical losses during one round trip, and the phase modulator is added to control the asymmetric phase distribution of the optical pulse. More importantly, since the phase on the left side of the mesh lattice with $n \leq 0$ is flexibly regulated, phase circumfluence is achieved at the position of $n=0$, so a binary synthetic photonic lattice with an effective magnetic field is constructed.
	
	In summary, this paper studies, theoretically and numerically, the reflection and transmission of light waves in a binary synthetic photonic lattice. By introducing asymmetric phase distribution in the lattices, effective magnetic circulation for photons can be obtained on one side of the interface. Such a tunable effective magnetic potential leads to both the reflection and energy exchange between two loops at the interface. Besides, the propagation constant $\Omega$ and wave vector $k$ also have important impacts on the light dynamics.  Especially, the light will be reflected totally and exchanges its energy completely in two loops at the interface --- anomalous reflection,  when $(\Phi_0,\Omega,k)=(\pi,\pi/2, n\pi)$. Our research proposes a new phenomenon by which it would be possible to achieve novel optical switches simply by fiber loops and phase modulators.	
	
	\section{Funding Information}
	This work is supported by the Open fund of Shandong Provincial Key Laboratory of Optics and Photonic Devices(K202008), Open Research Fund of State Key Laboratory of laser Interaction with Matter(SKLLIM1812), and Major Research plan of the National Natural Science Foundation of China(91950104).

	\section{Disclosures}
	The authors declare no conflicts of interest. And special thanks to Prof. Xin Guoguo for the constructive discussion.

	\bibliography{ref}
	
	\bibliographyfullrefs{ref}
	
	
	\ifthenelse{\equal{\journalref}{aop}}{%
		\section*{Author Biographies}
		\begingroup
		\setlength\intextsep{0pt}
		\begin{minipage}[t][6.3cm][t]{1.0\textwidth} 
			\begin{wrapfigure}{L}{0.25\textwidth}
				\includegraphics[width=0.25\textwidth]{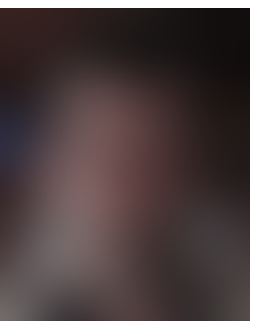}
			\end{wrapfigure}
			\noindent
			{\bfseries John Smith} received his BSc (Mathematics) in 2000 from The University of Maryland. His research interests include lasers and optics.
		\end{minipage}
		\begin{minipage}{1.0\textwidth}
			\begin{wrapfigure}{L}{0.25\textwidth}
				\includegraphics[width=0.25\textwidth]{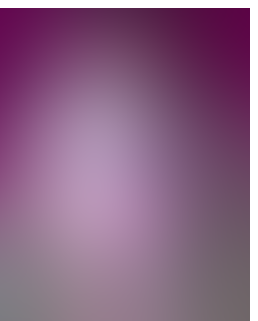}
			\end{wrapfigure}
			\noindent
			{\bfseries Alice Smith} also received her BSc (Mathematics) in 2000 from The University of Maryland. Her research interests also include lasers and optics.
		\end{minipage}
		\endgroup
	}{}

\end{document}